\documentclass[preprint,showpacs,preprintnumbers,amsmath,amssymb,superscriptaddress]{revtex4}

\usepackage{graphicx}
\usepackage{dcolumn}
\usepackage{bm}
\usepackage[usenames]{color}    

\usepackage{graphicx}
\usepackage{bm}
\usepackage{color}

\newcommand{\be}{\begin{equation}}
\newcommand{\ee}{\end{equation}}
\newcommand{\bea}{\vspace{0.25cm}\begin{eqnarray}}
\newcommand{\eea}{\end{eqnarray}}

\def\Int{{\cal I}}
\def\hN{{\hat{N}}}

\begin{document}
\title{Testing Quantum Gravity by Quantum Light}

\author{I.~Ruo Berchera}
\affiliation{INRIM, Strada delle Cacce 91, I-10135 Torino, Italy}

\author{I.~P.~Degiovanni}
\affiliation{INRIM, Strada delle Cacce 91, I-10135 Torino, Italy}

\author{S.~Olivares}
\affiliation{Dipartimento di Fisica, Universit\`a degli Studi di Milano, and CNISM UdR Milano Statale, Via Celoria 16, I-20133 Milano, Italy}

\author{M.~Genovese}
\affiliation{INRIM, Strada delle Cacce 91, I-10135 Torino, Italy}

\begin{abstract}
In the last years quantum correlations received large attention as key ingredient in advanced quantum metrology protocols, in this letter we show that they provide even larger advantages when considering multiple-interferometer setups.
In particular we demonstrate that the use of quantum correlated light
beams in coupled interferometers leads to substantial advantages with
respect to classical light, up to a noise-free scenario for the ideal
lossless case.
On the one hand, our results prompt the possibility of testing
quantum gravity in experimental configurations affordable in current
quantum optics laboratories and strongly improve the precision in "larger size experiments" such as the Fermilab holometer; on the other hand, they pave the way for future applications to high precision measurements and quantum metrology
\end{abstract}
\maketitle


The dream of building a theory unifying general relativity and quantum mechanics, the so called quantum gravity (QG), has been a key element in theoretical physics research for
the last 60 years.  Several attempts in this sense have been considered. However, for many years no testable prediction emerged from these studies, leading to the common wisdom that this kind of research was more properly a part of mathematics than of physics, being by construction unable to produce experimentally testable predictions as required by Galilean scientific method.
In the last few years this common wisdom was challenged \cite{am1,am2,am3,hog,alt,bek:12}. 
More recently, effects in interferometers  connected to non-commutativity of position variables \cite{ac,ac2} in different directions have been considered both for cavities with microresonators \cite{alt} and two coupled interferometers \cite{hog}, the so called ``holometer''. In particular this last idea led to the planning of a double 40~m interferometer at Fermilab \cite{www.holometer}.
\par
Here we consider whether the use of quantum correlated light beams in coupled interferometers could lead to significant improvements allowing an actual simplification of the experimental apparatuses to probe the non-commutativity of position variables.  On the one hand, our results demonstrate that the use of quantum correlated light can lead to substantial advantages in interferometric schemes also in the presence of non-unit quantum efficiency, up to a noise-free scenario for the ideal lossless case. This represents a big step forward respect to the quantum metrology schemes reported in literature \cite{k,giov:11,Huelga:97,Lee:02}, and paves the way for future metrology applications. On the other hand, they prompt the possibility of testing QG in experimental configurations affordable in a traditional quantum optics laboratory with current technology.
\par
The idea at the basis of the holometer is that non-commutativity at the Planck scale ($l_p = 1.616 \times 10^{-35}$ m) of position variables in different directions  leads
to an additional phase noise, referred to as holographic noise (HN). In a single interferometer $\Int$ this noise substantially confounds with other sources of noise,
even though the most sensible gravitational wave interferometers are considered \cite{hog}, since their HN resolution is worse than their resolution to gravitational-wave at low frequencies. Nonetheless, if the two equal interferometers $\Int_1$ and $\Int_2$ of the holometer have overlapping spacetime volumes, then the HN between them is correlated and easier to be identified \cite{hog}. Indeed, the ultimate limit for holometer sensibility, as for any classical-light based apparatus, is dictated by the shot noise: therefore, the possibility of going beyond this limit by exploiting quantum optical states is of the greatest interest \cite{giov:11,sch:10,bri:10}.
\par
In the past the possibility of exceeding shot-noise limit in gravitational-wave detectors was suggested \cite{cav,mc} and, recently, realized \cite{ab} by using squeezed light. As shown in the following, this resource can indeed allow an improvement of holometer-like apparatuses as well. Nonetheless, in this case, having two coupled interferometers, the full exploitation of properties of quantum light, and in particular of entanglement, may lead to much larger improvements.
\par
In general, the observable measured at the output of the holometer may be described by an appropriate operator $\widehat{C}(\phi_1, \phi_2)$, $\phi_k$ being the phase shift (PS) detected by $\Int_k$, $k=1,2$, with expectation value $\langle\widehat{C}(\phi_{1}, \phi_{2}) \rangle=\mathrm{Tr}[\rho_{12}\widehat{C}(\phi_{1}, \phi_{2}) ]$, where
$\rho_{12}$ is the overall density matrix associated with the state of the light beams injected in $\Int_1$ and $\Int_2$.
\par

\begin{figure}[tbp] \begin{center}
\includegraphics[width=0.4\textwidth]{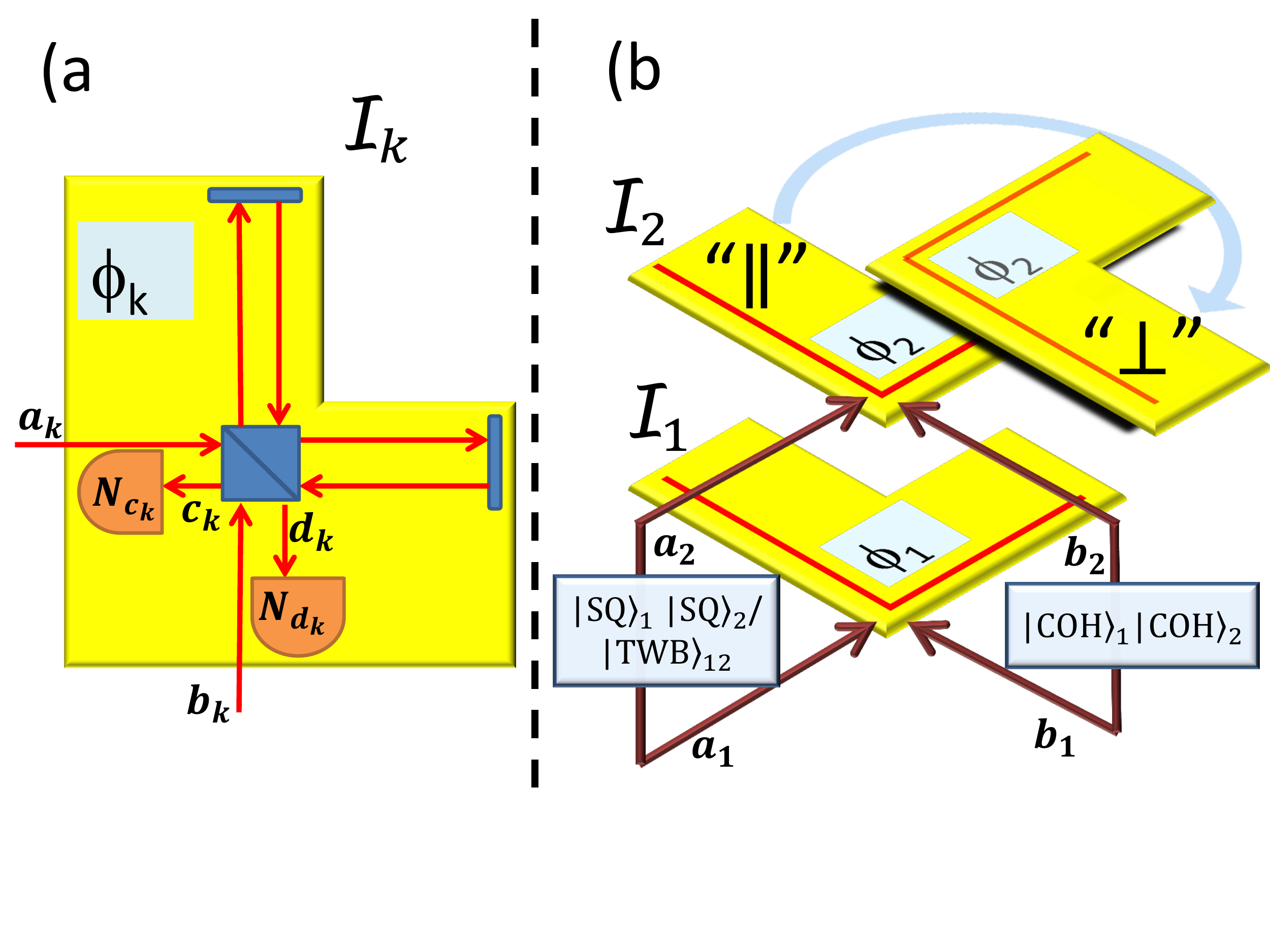}
\caption{Sketch of the holometer. a) The two involved interferometers, $\Int_k$, $k=1,2$, have input modes $a_k$ and $b_k$ and output modes $c_k\equiv c_k(\phi_k)$ and $d_k  \equiv d_k(\phi_k)$, where two detectors are placed for measuring the number of photons $\hN_{c_k}(\phi_k)$ and $\hN_{d_k}(\phi_k)$, respectively. b) The interferometers are set in the configuration ``$\parallel$"  and ``$\perp$'' according to the picture. The input modes $b_k$, $k=1,2$, are always excited in two coherent states labeled as "$|COH\rangle_{1}|COH\rangle_{2}$", while the modes $a_k$ are excited in two uncorrelated squeezed vacua labeled as "$|SQ\rangle_{1}|SQ\rangle_{2}$" [configuration (SQ)], or in a maximally entangled two-mode squeezed vacuum marked as "$|TWB\rangle_{12}"$ [configuration (TWB)]. }.
\label{Fig1} \end{center} \end{figure}

In order to observe the HN, one should compare \cite{hog}  $\langle\widehat{C}(\phi_{1}, \phi_{2}) \rangle$ in two different experimental configurations of $\Int_1$ and $\Int_2$, namely,
parallel, ``$\parallel$'', and perpendicular, ``$\perp$'' (Fig. \ref{Fig1}). In configuration``$\parallel$'', the interferometers are oriented so that the HN induces the same random fluctuation of their PSs, leading to a substantial correlation between them, since they occupy overlapping space-time volumes \cite{hog, footnoteA}. Thus, by measuring the correlation of the interference fringes, one can highlight the presence of the HN. Configuration ``$\perp$'' serves as a reference measurement, namely, it corresponds to the situation where the correlation due to HN is absent, since their space-time volumes are not overlapping \cite{hog, footnoteA}, in other words, it is equivalent to the estimation of the ``background''.  The statistical properties of the PSs fluctuations due to HN may be described by a suitable probability density function, $f_x (\phi_{1}, \phi_{2})$, $x=\parallel,\perp$. In turn, the expectation of any operator $\widehat{O}(\phi_1,\phi_2)$, or function of the PSs, should be averaged over $f_x$, namely,
\begin{equation}\langle\widehat{O}(\phi_1,\phi_2)\rangle \to
\mathcal{E}_{x}\left[\widehat{O}(\phi_1,\phi_2)\right]
\equiv \int \langle\widehat{O}(\phi_1,\phi_2)\rangle ~ f_x (\phi_{1}, \phi_{2})  ~
\mathrm{d} \phi_{1} ~ \mathrm{d} \phi_{2}.\label{f-average}
\end{equation}
\par
As in the holometer the HN arises as a correlation between two phases, the appropriate function to be estimated is their covariance in the
parallel configuration $\mathcal{E}_\parallel\left [ \delta \phi_{1} \delta \phi_{2}\right]$, where $\delta \phi_{k} = \phi_k - \phi_{k,0}$, and $\phi_{k,0}$
are the mean PS value measured by $\Int_k$, $k=1,2$. Since the holographic noise is supposed to be small, we can expand the $\widehat{C}$ operator in terms of small fluctuation
$\delta \phi_{k}$. According to Eq. (\ref{f-average}) we are able to directly relate the covariance of the PSs to the observable quantities (see Sup.~Mat.~Sec.~I for details):
\begin{equation}\label{cov:PS}
\mathcal{E}_\parallel\left [ \delta \phi_{1} \delta \phi_{2}\right] \approx
\frac{
\mathcal{E}_\parallel \left[ \widehat{C}(\phi_1,\phi_2)\right]-
\mathcal{E}_\perp \left[ \widehat{C}(\phi_1,\phi_2)\right]}
{\langle \partial_{\phi_{1},\phi_{2}}^{2}
\widehat{C}(\phi_{1,0}, \phi_{2,0}) \rangle }, \quad
(\delta \phi_{1},\delta\phi_{2} \ll 1)
\end{equation}
Eq. (\ref{cov:PS}) states that the covariance can be estimated by measuring the difference between the expectation value of the operator $\widehat{C}$ in the two configurations.
Thus, this difference represents the measured signal, while the coefficient at the denominator is the sensitivity.

One has to reduce as much as possible the uncertainty associated with its measurement:
\begin{equation}
 \mathcal{U}(\delta \phi_{1} \delta \phi_{2}) \approx
\sqrt{\frac{\mathrm{Var}_\parallel \left[ \widehat{C}(\phi_1,\phi_2)
 \right] + \mathrm{Var}_\perp  \left[ \widehat{C}(\phi_1,\phi_2) \right]}{\left[ \langle
\partial_{\phi_{1},\phi_{2}}^{2}  \widehat{C}(\phi_{1,0}, \phi_{2,0}) \rangle \right]^2 }}, \quad
(\delta \phi_{1},\delta\phi_{2} \ll 1)
\end{equation}\label{U}
where $\mathrm{Var}_x \left[ \widehat{C}(\phi_1,\phi_2) \right] \equiv
\mathcal{E}_x \left[ \widehat{C}^2 (\phi_1,\phi_2) \right] -
\mathcal{E}_x \left[ \widehat{C}(\phi_1,\phi_2) \right]^2$ \cite{footnote1}. We observe
that the sum of variances derives from the independence of the two measurements configurations.
Thanks to the same expansions leading to Eq.~(\ref{cov:PS}), we can write
$\mathrm{Var}_x \left[ \widehat{C}(\phi_1,\phi_2) \right] = \mathrm{Var}\left[ \widehat{C}(\phi_{1,0 }, \phi_{2,0})
\right]+ \mathcal{O}(\delta \phi^2) $ for both $x="\parallel","\perp"$. Therefore, the zero-order contribution to the uncertainty is
\begin{equation} \label{U0} \mathcal{U}^{(0)} = \frac{\sqrt{2\, \mathrm{Var}\left[ \widehat{C}(\phi_{1,0 }, \phi_{2,0}) \right]}}{\left|
 \langle \partial_{\phi_{1},\phi_{2}}^{2} \widehat{C}(\phi_{1,0},
 \phi_{2,0}) \rangle \right|},
\end{equation}
where $\mathrm{Var}\left[\widehat{C}(\phi_{1,0 }, \phi_{2,0}) \right]=
\langle
\widehat{C}(\phi_{1,0}, \phi_{2,0})^2 \rangle - \langle
\widehat{C}(\phi_{1,0 }, \phi_{2,0}) \rangle^2 $ does not depend on the
PSs fluctuations due to the HN, but it represents the intrinsic
quantum fluctuations of the measurement described by the operator
$\widehat{C}(\phi_1,\phi_2)$ and depends on the optical quantum states
sent in the holometer. In particular, our aim is to look for a suitable choice of quantum optical state $\rho_{12}$ and an operator $\widehat{C}(\phi_1,\phi_2)$ that reduces this zero-order contribution to the uncertainty.
In the following we will demonstrate that the use of quantum resources, like squeezing or, much more, entanglement, provides huge advantages in terms of achieved accuracy with respect to classical light.
\par
As a first example we consider a configuration (SQ) where the two input modes of each interferometer $\Int_k$, $k=1,2$, are excited in a coherent state and a squeezed vacuum state with mean number of photons $\mu_k$ and $\lambda_k$, respectively (see Fig.~\ref{Fig1}). Since the difference of the number of photons in the two output ports of each interferometer, $\hN_{k-}(\phi_k) = \hN_{c_k}(\phi_k)-\hN_{d_k}(\phi_k)$, can be used to estimate the corresponding $\phi_{k}$ with sub-shot-noise resolution \cite{k,giov:11,cav}, then  reasonably the covariance $\mathcal{E}_\parallel [ \delta \phi_{1}\delta \phi_{2}]$ can be efficiently evaluated from the covariance between $\hN_{1-}(\phi_1)$ and $\hN_{2-}(\phi_2)$. Therefore we define $\widehat{C}(\phi_1,\phi_2) = \Delta \hN_{1-}(\phi_1) ~ \Delta \hN_{2-}(\phi_2)$, with $\Delta \hN_{k-}(\phi_k)= \hN_{k-}(\phi_k)- \mathcal{E}\left[ \hN_{k-}(\phi_k)\right]$ (we note that $\mathcal{E}_\parallel\left[ \hN_{k-}(\phi_k)\right]=\mathcal{E}_\perp\left[ \hN_{k-}(\phi_k)\right]= \mathcal{E}\left[ \hN_{k-}(\phi_k)\right]$, as consequence of the properties of $f_x (\phi_{1}, \phi_{2})$, see Sup.~Mat.~Sec.~I).

\begin{figure}[tbp]\begin{center}
\includegraphics[width=0.4\textwidth]{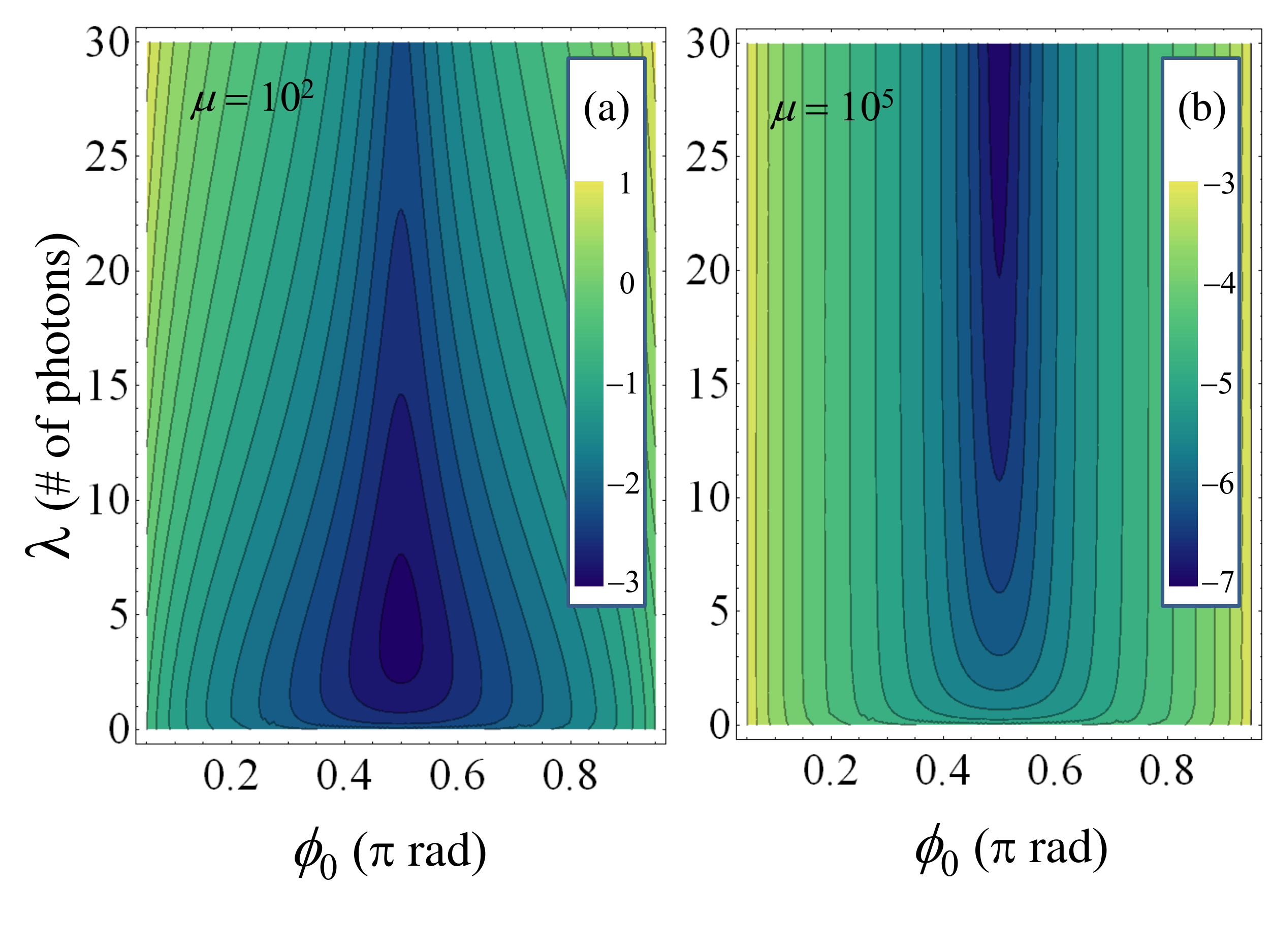}
\caption{The uncertainty of the phase covariance $\log_{10}\mathcal{U}_{SQ}^{(0)}$ when the holometer is fed by two independent squeezed states plus two coherent fields. Here $\phi_{1,0}=\phi_{2,0}=\phi_{0}$ is the central PS of the interferometers, $\lambda_{1}=\lambda_{2}=\lambda$ is the intensity of each squeezing, $\mu_{1}=\mu_{2}=\mu$ is the intensity of each coherent beam, and their phase difference is set to zero. (a) $\mu \sim \lambda$: when the coherent and squeezed beams have similar intensities the noise reduction is lower bounded. (b) $\mu \gg \lambda$:  in this regime quantum noise reduction increases with the amount of squeezing, offering a strong noise suppression if high level of squeezing can be reached. The region of the minimum runs at $\phi_{0}=\pi/2$.} \label{fig_U0s} \end{center} \end{figure}

Fig.~\ref{fig_U0s} shows the corresponding uncertainty at the zero-order given in Eq.~(\ref{U0}): assuming identical input states ($\mu_k=\mu$ and $\lambda_k=\lambda$, $k=1,2$), the minimum is achieved for $\phi_{1,0}=\phi_{2,0}= \pi/2$, and reads (see Sup.~Mat.~Sec.~II):
\begin{equation}\label{U0sq}
\mathcal{U}_{\rm SQ}^{(0)} (\mu,\lambda)\approx\sqrt{2}~\frac{\lambda + \mu \left(1+2 \lambda - 2 \sqrt{\lambda
+\lambda^2}\right)}{(\lambda-\mu)^2}.
\end{equation}
In perfect analogy with the PS measurement for a single interferometer \cite{cav,mc}, if $\mu \gg \lambda \gg 1$, then we have the optimal accuracy $\mathcal{U}_{\rm SQ}^{(0)} \approx (2\sqrt{2} \lambda \mu )^{-1}$. This represents a strong advantage in terms of uncertainty reduction [of the order $(4 \lambda )^{-1}$] with respect to classical case $\mathcal{U}_{\rm CL}^{(0)}\approx \sqrt{2}/ \mu $, {\it i.e.} when only coherent states are employed.
Nevertheless, an important difference between the single interferometer PS measurement, involving a first order moment of the photon number distribution,  and the covariance estimation, involving the second order momenta, arises: while in the first case the uncertainty scales as the usual standard quantum limit one, $\propto\mu^{-1/2}$, in the second case it scales  $\propto\mu^{-1}$ (neglecting the little relative contribution of the squeezing to the intensity). We remark that the advantage of the present scheme is based on the independent improvement of the resolution of each interferometer which is itself limited by the amount of squeezing (see Sup.~Mat.~Sec.~II). However, the aim of the holometer is to couple $\Int_1$ and $\Int_2$ minimizing the noise on their outputs correlation, namely regardless of the noise in the single interferometer. This suggests that quantum correlated states, coupling $\Int_1$ and $\Int_2$, could further enhance the performance of the holometer.
\par
To this aim, we consider a new configuration (TWB) where modes $a_1$ and $a_2$ of Fig.\ref{Fig1} are excited in a
continuous variable maximally entangled state, i.e. a two-mode squeezed vacuum
state or twin-beam state, $|{\rm TWB}\rangle\rangle_{a_1,a_2} =
S_{12}(\zeta) |0\rangle_{a_1,a_2} = \sum_{n=0}^{\infty} c_n(\zeta)
|n\rangle_{a_1}|n\rangle_{a_2}$, where $S_{12}(\zeta)=\exp(\zeta
~a_{1}^\dag a_{2}^\dag - \zeta^* ~a_{1} a_{2})$ is the two-mode
squeezing operator. This state can be easily produced experimentally, for example by the parametric-down conversion process  \cite{MandelWolf}. If we set $\zeta= |\zeta| e^{i\theta_{\zeta}}$ and
introduce the mean photon number per mode $\lambda=\sinh^2 |\zeta|$,
then $c_n(\zeta) =
(1+\lambda)^{-1/2}\left[(1+\lambda^{-1})~e^{-i2\theta_{\zeta}}\right]^{-n/2}$
\cite{oli:12}. The input modes $b_1$ and $b_2$ are still excited in two coherent states, so that the four-mode input state is $|\psi\rangle=|{\rm
  TWB}\rangle\rangle_{a_1,a_2}\otimes|\alpha\rangle_{b_1}\otimes|\alpha\rangle_{b_2}$.

One of the peculiarity of the state $|{\rm TWB}\rangle\rangle_{a_1,a_2}$ is the presence of the same number of photons in the two modes \cite{bri:10,pra,mas,m}, then each power of the photon number difference of the two modes is identically null, ${}_{a_1,a_2}\langle\langle{\rm TWB}|(\hat{N}_{a_1}-\hat{N}_{a_2})^{M}|{\rm
TWB}\rangle\rangle_{a_1,a_2}=0$, $\forall M>0$.

We also observe that, in the absence of the HN and choosing the optimal working regime $\phi_{k,0}=0$, $k=1,2$, $\Int_1$ and $\Int_2$ behave
like two completely transparent media for their input fields [see
again Sup. Mat. Sec. II, Eq.s (4)]. In particular output modes $c_1$ and $c_2$ exhibit perfect correlation between the number of photons, which directly comes from  the input modes $a_1$ and $a_2$, leading to the the natural choice of the observable as the fluctuation of the photon numbers difference, $\widehat{C}(\phi_1,\phi_2)=\Delta^{2}\left[\hN_{c_1}-\hN_{c_2}\right]$. Indeed, the numerator of Eq.~(\ref{U0}),
$\mathrm{Var}\left\{ \Delta^{2}\left[\hN_{c_1}(0)-\hN_{c_2}(0)\right]\right\}$ is identically null, while the denominator reads:
\begin{equation}\label{K}
  \left\langle\psi\left|\partial_{\phi_{1},\phi_{2}}^{2}
      \Delta^2\left[\hN_{c_1}(\phi_{1,0})-\hN_{c_2}(\phi_{2,0})\right]\right|\psi\right\rangle=
-\frac{1}{2} \sqrt{\lambda (1 + \lambda)}
  \mu \cos[2 (\theta_{\zeta}-\theta_{\alpha})]
\end{equation}
that is non-zero for both $\lambda,\mu\ne 0$ and it is maximized for
$\theta_{\zeta}-\theta_{\alpha}=0$.
This quantity represents also the coefficient of proportionality in Eq.~(\ref{cov:PS}) between the covariance of the HN and the measured signal.
It is worth noting that, even though for $\phi_{k,0}=0$ the coherent state
gives no contribution to the output modes $c_1$ and $c_2$, being
completely transmitted to the complementary modes $d_1$ and $d_2$, when
fluctuations of the PS occurs a little portion will be reflected to
the monitored ports and this guarantees the sensitivity to the HN PSs
covariance.

Thus, the correlation property of the TWB state leads to the amazing result that the contribution to the uncertainty coming from the photon number noise shown in Eq.~(\ref{U0}) is $\mathcal{U}_{\rm TWB}^{(0)}=0$ (when $\lambda,\mu\ne0$), representing an ideal accuracy of the interferometric scheme to the PSs covariance due to HN and the main achievement of the present study.

\par

\begin{figure}[htbp]
\begin{center}
\includegraphics[width=0.8\textwidth]{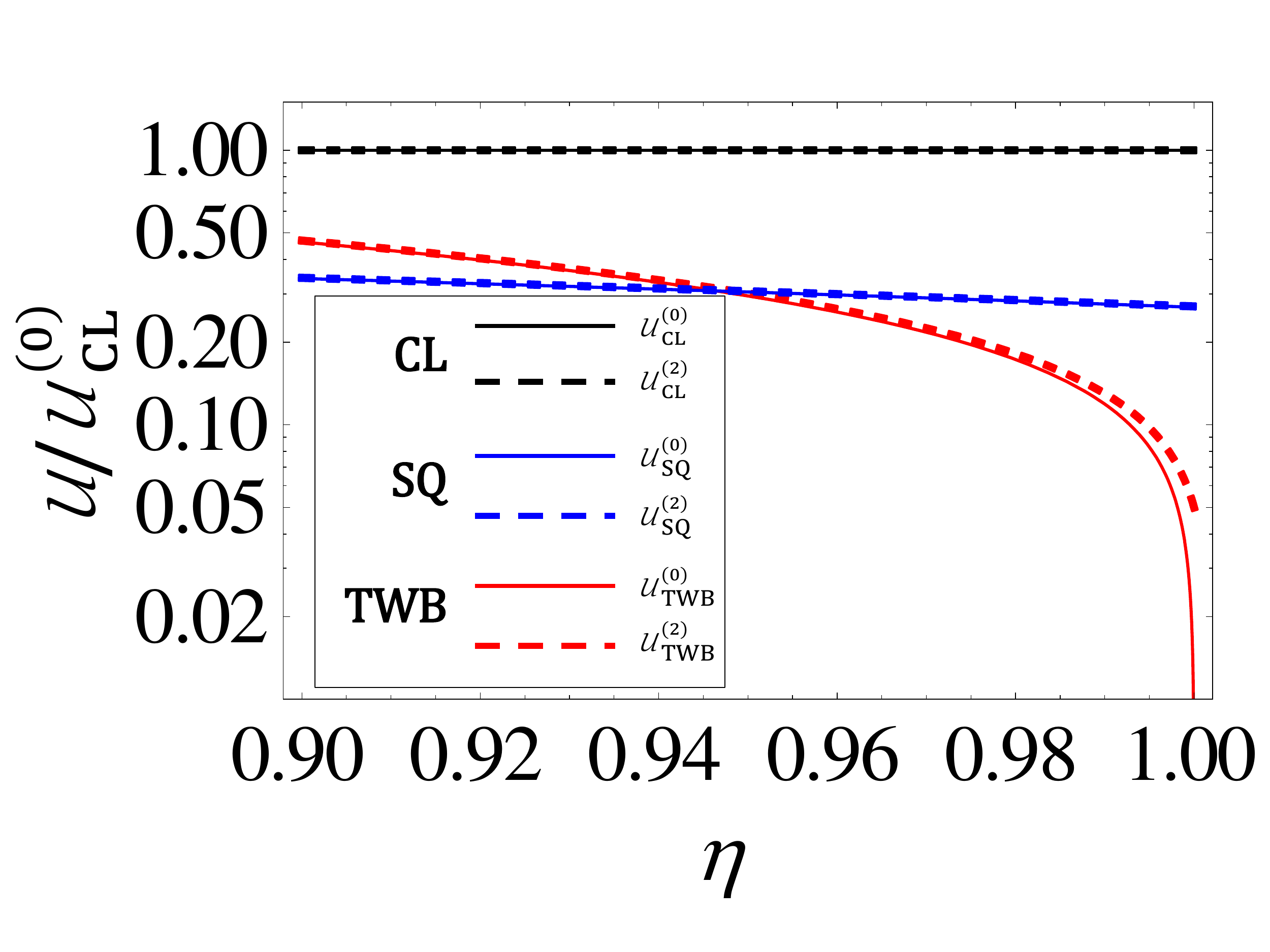}
\caption{Uncertainty reduction normalized to the classical limit $\mathcal{U}_{CL}^{(0)}$ \cite{footnote2}. The solid lines represent the uncertainties only due to the photon noise,
corresponding to the zero-order contribution [see Eq.~(\ref{U0})]. The coherent field intensity is set to $\mu=2\times 10^{23}$, while the twin beams and independent
squeezed beams intensities are $\lambda_{1}=\lambda_{2}=\lambda=0.5$. The dashed lines represent the second-order uncertainties including the RP contribution. For its calculation the measurement time is chosen $\tau=10^{-3}$~s, the mirror mass $m=10^{2}$~Kg, and the central angular frequency of the light $\omega=3.14\times 10^{15}$~Hz (a wavelength of 600 nm.)}
\label{fig_RP_eta} \end{center} \end{figure}

The question that now arises is how and at which extent our conclusions are affected by a non unity overall transmission-detection efficiency $\eta$ (see Sup.~Mat.~Sec.~III).
In Fig.~\ref{fig_RP_eta} we plot $\mathcal{U}^{(0)}$ for the SQ, TWB and classical coherent state (CL) approaches, as a function of $\eta$ (assumed to be the same for both the interferometers) for a modest level of non-classical resources ($\lambda=0.5$). As one may expect, SQ exhibits a little advantage in this regime. However, in the high efficiency region (albeit with values reasonable achievable with current technologies)  the TWB-based approach provides a significant improvement not only with respect to classical set-up, but also respect to SQ.
\par
Focusing on the limit of very small quantum resources, {\it i.e.} $\lambda\ll1$ and $\mu\gg1$, then $\mathcal{U}_{\rm SQ}^{(0)}/\mathcal{U}_{\rm CL}^{(0)} \approx  1-2\eta\sqrt{\lambda}$ and $\mathcal{U}_{\rm TWB}^{(0)}/\mathcal{U}_{\rm CL}^{(0)} \approx \sqrt{2(1-\eta)/\eta}$. For a small amount of squeezing the quantum noise $\mathcal{U}_{\rm SQ}^{(0)}$ not surprisingly approaches the classical case, while for TWB, we have a degradation of the performances with respect to the ideal case ($\eta=1$). Anyway, an improvement  with respect to the classical case is kept for $\eta>2/3$, demonstrating that a  relatively faint TWB can provide an interesting improvement in the HN detection.
\par
In the opposite limit of high quantum resources exploited, {\it i.e.} $\mu\gg\lambda\gg1$,
$\mathcal{U}_{\rm SQ}^{(0)}/\mathcal{U}_{\rm CL}^{(0)} \approx \left( 1-\eta\right)+\eta/(4\lambda)$ and
$\mathcal{U}_{\rm TWB}^{(0)}/\mathcal{U}_{\rm CL}^{(0)} \approx 2\sqrt{5}\left( 1-\eta\right)$
reveal that the performances of the quantum strategies
are limited by  the presence of the terms $(1-\eta)$.  Here the main difference between SQ and TWB is that for $\eta\approx1$ SQ exhibits an uncertainty lower bounded to $(4\lambda)^{-1}$, {\it i.e.} depending on the squeezing intensity. On the other hand, TWB approach beats the classical one for $\eta>0.683$, while for $\eta\approx 1$ it goes to zero: this demonstrates that the use of quantum light can largely improve the performances of a holometer addressed to test quantum gravity models.
\par

A last source of noise, which could affect our results, may derive from the radiation pressure (RP) \cite{Braginsky,cav,Kimble}.
However, our model is perturbative in phase fluctuations $\delta \phi_{k}$ and,
according to Eq.~(\ref{U0sq}), the main noise contribution should come
from the zeroth-order term corresponding to the photon noise.
RP is assumed to introduce
a second-order contribution to the uncertainty [see Sup. Mat. Sec. IV for details], that is related to the light fluctuation
in the arms of the interferometers and to the phase shift induced by the mirrors recoil
(the latter is given by  $(\omega \tau/2 m c)\mathcal{P}$, where $\omega$ and $\mathcal{P}$ are
the central frequency of the light and the momentum of the photon, respectively,
$\tau$ the measurement time and $m$ the mirrors mass \cite{cav}).
In the case of a single interferometer fed by squeezed light, the amplitude of the RP noise
varies with the squeezing parameter at the opposite of the photon noise, namely, if one decreases the other increases and viceversa, thus an optimum regime must be found.
In the context of our proposal the behavior is similar, but for reasonable values of the involved
parameters RP noise is completely negligible (justifying our perturbative approach) .  Fig.~\ref{fig_RP_eta} shows how RP noise starts to affect the uncertainty for an average coherent field intensity $\mu \gtrsim 10^{23}$ for $\tau=10^{-3}$~s (the photons introduced by the quantum modes are negligible), that is a power larger than $\hbar\omega\mu/\tau=3.3\times 10^{7}$~Watts! Since the HN must be seek in the region of the MHz, i.e. for a short measurement time ($\tau \approx 10^{-6}$~s), the RP contribution would be significant for power values larger than $10^{13}$~Watts, a value well far away from the current and future interferometry technology. One can be surprised that in the present scheme the radiation pressure is negligible, while it is not always the case for the standard phase measurement involving a single interferometer. However, we stress that here we are measuring a phase covariance between two interferometers, instead of the phase values in single ones.

In conclusion, besides our analysis concerning the use of two independent squeezed states, which substantially confirms the advantages already demonstrated for gravity waves detection, the investigation carried out exploiting entanglement leads to the unprecedented result that, in an ideal situation, the background noise can be completely washed out. This achievement not only paves the way for reaching much higher sensibility in the holometer in construction at Fermilab or for the realisation of a table top experiment to test quantum gravity, but also sheds some first light on new unexpected opportunities offered by the use quantum states of light for a fundamental reduction of noise in interferometric schemes.

\section{Supplementary material}

\subsection{Derivation of Eq.~(2) of the main text}

Since the fluctuations due to the holographic
noise (HN) are expected to be extremely small, we can expand $\widehat{C}(\phi_1, \phi_2)$ around the phase shift (PS) central
values $\phi_{1,0}$, $\phi_{2,0}$, namely:
\begin{eqnarray}
\widehat{C}(\phi_1, \phi_2)&=&\widehat{C}(\phi_{1,0}, \phi_{2,0})+ \Sigma_k ~ \partial_{\phi_{k}}
\widehat{C}(\phi_{1,0}, \phi_{2,0})~ \delta \phi_k + \frac{1}{2} \Sigma_k ~
\partial_{\phi_{k},\phi_{k}}^{2} \widehat{C}(\phi_{1,0}, \phi_{2,0})~ \delta \phi_{k}^2 \nonumber \\
&\hbox{}&+
\partial_{\phi_{1},\phi_{2}}^{2} \widehat{C}(\phi_{1,0}, \phi_{2,0}) ~ \delta \phi_{1} \delta
\phi_{2}+ \mathcal{O}(\delta \phi^3)
\label{C}
\end{eqnarray}
where  $\delta \phi_{k}= \phi_{k}-\phi_{k,0}$, and
$\partial_{\phi_{1}^h, \phi_{2}^k}^{h+k}\widehat{C}(\phi_{1,0},
\phi_{2,0}) $ is the $(h+k)$-th order derivative of
$\widehat{C}(\phi_{1}, \phi_{2}) $ calculated  at $\phi_{k}=\phi_{k,0}$, $k,h=1,2$
\par
In order to reveal the HN, the holometer exploits two different
configurations: the one, ``$\parallel$", where HN correlates the
interferometers, the other, ``$\perp$ ", where the effect of HN
vanishes (Fig.~(1) of the main text).  The statistical properties of the PS
fluctuations due to the HN may be described by the joint probability
density functions $f_\parallel (\phi_{1}, \phi_{2})$ and $f_\perp
(\phi_{1}, \phi_{2})$. We make two reasonable hypotheses about
$f_x (\phi_{1}, \phi_{2})$, $x=\parallel,\perp$.  First, the marginals
$\mathcal{F}^{(k)}_{x} (\phi_{k})=\int \mathrm{d} \phi_{h } f_x
(\phi_{1}, \phi_{2})$, $h,k=1,2$ with $h\neq k$, are exactly the same
in the two configurations, i.e.
$\mathcal{F}^{(k)}_{\parallel}(\phi_{k})=\mathcal{F}^{(k)}_{\perp}(\phi_{k})$:
one cannot distinguish between the two configurations just by
addressing one interferometer. Second, only in configuration
``$\perp$'' we can write $f_\perp (\phi_{1}, \phi_{2})=
\mathcal{F}^{(1)}_{\perp}(\phi_{1})\mathcal{F}^{(2)}_{\perp} (\phi_{2})$,
i.e., there is no correlation between the PSs due to the HN.  Now, the
expectation of any operator $\widehat{O}(\phi_1,\phi_2)$ should be
averaged over $f_x$, namely, $\langle\widehat{O}(\phi_1,\phi_2)\rangle
\to \mathcal{E}_{x}\left[\widehat{O}(\phi_1,\phi_2)\right] \equiv \int
\langle\widehat{O}(\phi_1,\phi_2)\rangle ~ f_x (\phi_{1},
\phi_{2}) ~ \mathrm{d} \phi_{1} ~ \mathrm{d} \phi_{2}$.  In
turn, by averaging the expectation of Eq.~(\ref{C}), we have:
\begin{eqnarray}
\mathcal{E}_x \left[ \widehat{C}(\phi_1, \phi_2) \right]
&=&
\langle \widehat{C}(\phi_{1,0}, \phi_{2,0}) \rangle
+  \frac{1}{2} \Sigma_k ~
\langle \partial_{\phi_{k},\phi_{k}}^{2}  \widehat{C}(\phi_{1,0},
\phi_{2,0}) \rangle ~ \mathcal{E}_x \left[\delta \phi_{k}^2\right]
\nonumber \\
& \hbox{} & +
 \langle \partial_{\phi_{1},\phi_{2}}^{2}\widehat{C}(\phi_{1,0},
\phi_{2,0}) \rangle~  \mathcal{E}_x \left[ \delta \phi_{1} \delta
\phi_{2}\right]+ \mathcal{O}(\delta \phi^3)
 \label{EC}
\end{eqnarray}
where we used $\mathcal{E}_x \left[\delta
  \phi_k\right]=0$. Then, according to the assumption on $f_x (\phi_{1}, \phi_{2})$ we have $\mathcal{E}_\parallel \left[\delta
  \phi_{k}^2\right]= \mathcal{E}_\perp \left[\delta
  \phi_{k}^2\right]$ and $\mathcal{E}_\perp \left[\delta \phi_{1} \delta
\phi_{2}\right] =\mathcal{E}_\perp \left[\delta \phi_{1} \right] \mathcal{E}_\perp \left[\delta
\phi_{2}\right]= 0$, and from Eq.~(\ref{EC}) follows that the PS
covariance may be written as in Eq. (2) of the main text, namely:
\begin{equation}\label{PSs-Cov}
  \mathcal{E}_\parallel \left[ \delta \phi_{1} \delta \phi_{2} \right]
  \approx
  \frac{\mathcal{E}_\parallel \left[ \widehat{C}(\phi_1, \phi_2)
      \right]-\mathcal{E}_\perp \left[ \widehat{C}(\phi_1, \phi_2)\right]}{\langle \partial_{\phi_{1},\phi_{2}}^{2}\widehat{C}(\phi_{1,0},\phi_{2,0}) \rangle },
\end{equation}
that is proportional to the difference between the mean values of the operator
$\widehat{C}(\phi_1, \phi_2)$ as measured in the two configurations
``$\parallel$ and ``$\perp$''.

Indeed, one has to reduce as much as possible the uncertainty associated with its measurement, which reads as:
\begin{equation}\label{U}
 \mathcal{U}(\delta \phi_{1} \delta \phi_{2}) \approx
\sqrt{\frac{\mathrm{Var}_\parallel \left[ \widehat{C}(\phi_1,\phi_2)
 \right] + \mathrm{Var}_\perp  \left[ \widehat{C}(\phi_1,\phi_2) \right]}{\left[ \langle
\partial_{\phi_{1},\phi_{2}}^{2}  \widehat{C}(\phi_{1,0}, \phi_{2,0}) \rangle \right]^2 }}, \quad
(\delta \phi_{1},\delta\phi_{2} \ll 1)
\end{equation}
where $\mathrm{Var}_x \left[ \widehat{C}(\phi_1,\phi_2) \right] \equiv
\mathcal{E}_x \left[ \widehat{C}^2 (\phi_1,\phi_2) \right] -
\mathcal{E}_x \left[ \widehat{C}(\phi_1,\phi_2) \right]^2$.

Under the same hypotheses used for deriving Eq. (\ref{PSs-Cov}) we can calculate the variance of $\widehat{C}(\phi_1, \phi_2)$ as
\begin{eqnarray}\label{VarC}
\mathrm{Var}_x \left[ \widehat{C}(\phi_1,\phi_2) \right]=\mathrm{Var} \left[ \widehat{C}(\phi_{1,0},\phi_{2,0}) \right]+ \Sigma_k ~ A_{kk}~\mathcal{E}_x \left[\delta \phi_{k}^2\right]+ A_{12}~\mathcal{E}_x \left[\delta \phi_{1}\delta \phi_{2}\right]+\mathcal{O}(\delta \phi^3)
\end{eqnarray}
where:
\begin{eqnarray}\label{Akj}
A_{kk}&= &\langle \widehat{C}(\phi_{1,0},\phi_{2,0}) \partial_{\phi_{k},\phi_{k}}^{2}\widehat{C}(\phi_{1,0},\phi_{2,0})\rangle\\\nonumber
&+&\langle [\partial_{\phi_{k}} \widehat{C}(\phi_{1,0},\phi_{2,0})]^{2} \rangle-\langle \widehat{C}(\phi_{1,0},\phi_{2,0})\rangle\langle  \partial_{\phi_{k},\phi_{k}}^{2}\widehat{C}(\phi_{1,0},\phi_{2,0})\rangle\\\nonumber
A_{12}&=&2\langle \widehat{C}(\phi_{1,0},\phi_{2,0}) \partial_{\phi_{1},\phi_{2}}^{2}\widehat{C}(\phi_{1,0},\phi_{2,0})\rangle\\\nonumber
&+&2\langle \partial_{\phi_{1}} \widehat{C}(\phi_{1,0},\phi_{2,0})\partial_{\phi_{2}} \widehat{C}(\phi_{1,0},\phi_{2,0}) \rangle-\langle \widehat{C}(\phi_{1,0},\phi_{2,0})\rangle\langle  \partial_{\phi_{1},\phi_{2}}^{2}\widehat{C}(\phi_{1,0},\phi_{2,0})\rangle
\end{eqnarray}
Analyzing expression (\ref{VarC}), we note the presence of a
zeroth-order contribution that does not depend on the PSs intrinsic
fluctuations, and represents the quantum photon noise of the
measurement described by the operator $\widehat{C}(\phi_1,\phi_2)$
evaluated on the optical quantum states sent into the holometer. The
statistical characteristics of the phase noise enter as second-order
contributions in Eq.~(\ref{VarC}) from each interferometer plus a
contribution coming from phase correlation between them.

This work addresses specifically the problem of reducing the photon
noise below the shot noise in the measurement of the HN, therefore in
the following and in the main text starting from Eq.~(3), we assume
the zero-order contribution being the dominant one. Of course, this
means to look for the HN in a region of the noise spectrum that is
shot-noise limited. Since the HN is expected up to frequencies of tens
MHz, it follows that all the sources of mechanical vibration noise are
suppressed. The only additional source of noise could be the radiation
pressure, which is itself related to the light fluctuation in the arms
of the interferometers. Nevertheless, we will demonstrate in Sec.~\ref{Radiation
  pressure noise} that it is indeed negligible in our regime of interest.

\subsection{Elements for deriving Eq.~(5) of the main text}\label{Derivation of Eq.(3) of the main text}

\begin{figure}[tbp]
\begin{center}
\includegraphics[width=0.6\textwidth]{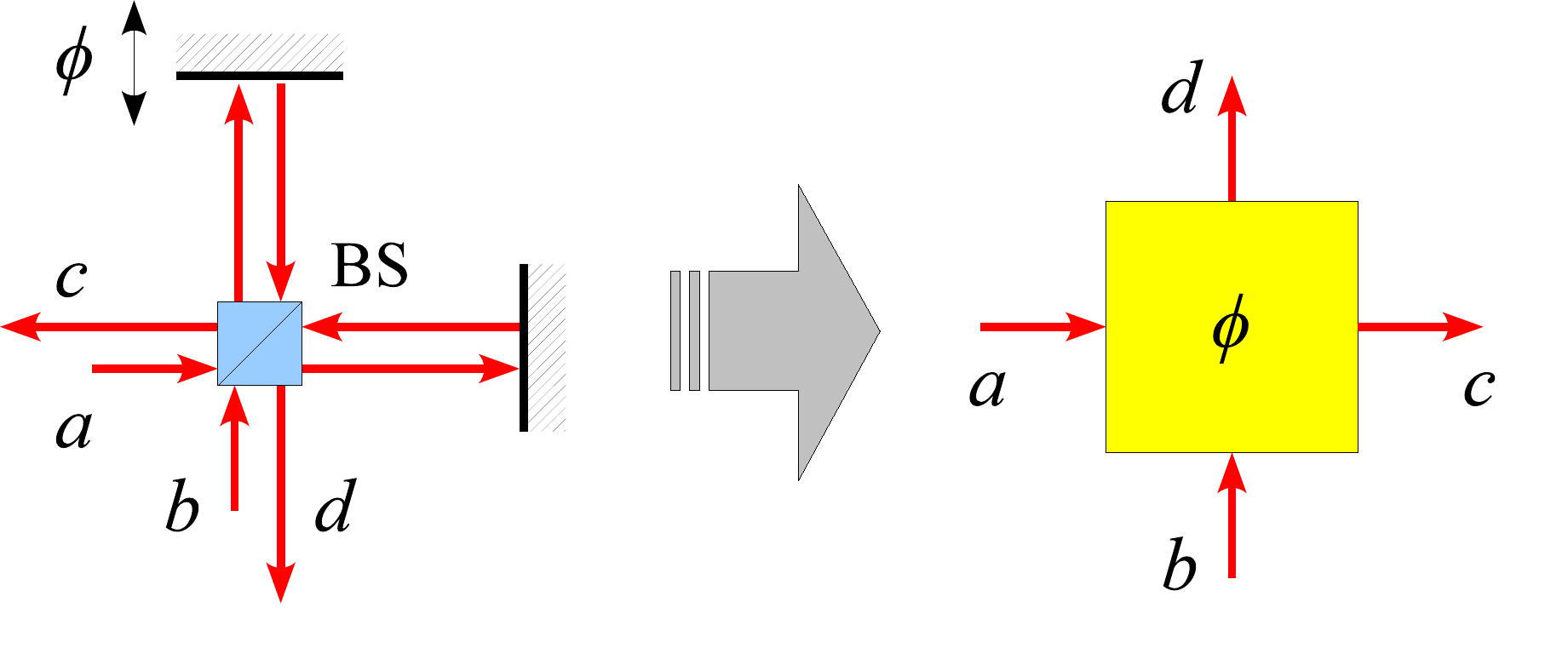}
\caption{The action of an optical interferometer, e.g., a Michelson
  interferometer (left), with input modes $a$ and $b$ and output modes
  $c$ and $d$ measuring an overall $\phi$ phase shift between the
  arms can be summarized as a beam-splitter like system with
  transmittance $\cos^2 (\phi/2) $ (right).  } \label{MtoS}
\end{center}
\end{figure}

At first, we briefly review a basic aspect of the quantum description
of optical interferometers. We consider two input modes described by
the bosonic field operators $a$ and $b$ ($[a,a^\dag]=1$ and
$[b,b^\dag]=1$), entering the two input ports of the optical
interferometer, for instance a Michelson interferometer, as depicted
in Fig.~\ref{MtoS}, whose main component is a 50:50 beam splitter
(BS). The two modes interfere a first time at the BS, then they are
reflected by two mirrors and interfere a second time at the BS after
having gained an overall $\phi$ phase shift) induced by the
difference between the optical paths of the two arms. We refer to the
outgoing modes as $c$ and $d$, respectively (Fig.~\ref{MtoS}).  The
input/output relations can be written as \cite{MandelWolf}:
\begin{equation}
\begin{array}{ll}\label{IOinterf}
c\equiv c(\phi)= \cos(\phi/2) ~  a -i \sin(\phi/2) ~  b , \\
d \equiv d(\phi)= -\cos(\phi/2) ~  b + i \sin(\phi/2)  ~ a .
\end{array}
\end{equation}
We note that such input/output relations are equivalent to those of
a BS with overall transmittance $\cos^2(\phi/2)$  (Fig.~\ref{MtoS}).

\par
It is well known, in quantum optics community, that coherent light
itself is not the optimal solution for PS estimation of an
interferometer, and that the use of squeezed light may enhance the
performance up to the Heisenberg limit \cite{pezze:08}.  In this
section we assume that the input modes $a_k$ and $b_k$ are excited in
a squeezed vacuum state $|\xi_k \rangle_{a_k} = S_{a_k}(\xi_k)|0
\rangle_{a_k}$ and the coherent state $| \alpha_k \rangle_{b_k} =
D_{b_k}(\alpha_k)|0 \rangle_{b_k}$, respectively, where
$S_{a_k}(\xi_k) = \exp[ \xi_k~(a_k^\dag)^2 - \xi_k^*~(a_k)^2]$ is the
squeezing operator and $D_{b_k}(\alpha_k) = \exp( \alpha_k~b_k^\dag -
\alpha_k^*~b_k)$ is the displacement operator. From now on, for the
sake of simplicity, we drop the subscript $k=1,2$. If we set
$\xi=|\xi| e^{i \theta_\xi}$ and $\alpha=\sqrt{\mu} e^{i
  \theta_\alpha}$, then $\lambda=\sinh^2 |\xi|$ and $\mu$ represent
the average number of photons of the squeezed vacuum and of the
coherent state, respectively. In the case of a single interferometer,
when PS is estimated from the measurement of the relative number of
photons $\hN_{-}(\phi)=\hN_c(\phi)-\hN_d(\phi)$ [with
$\hN_c(\phi)=c^\dag c$ and $\hN_d(\phi)=d^\dag d$, note that $c$ and
$d$ depends on $\phi$, see Eq.s~(\ref{IOinterf})], one has $\langle
\hN_{-}(\phi)\rangle=(\mu-\lambda)\cos\phi$, and in the limit
$\mu\gg\lambda$, around the optimal working point $\phi=\pi/2$, the
predicted uncertainty is $e^{-|\xi|}/\sqrt{\mu}$, i.e. below the shot
noise or standard quantum limit \cite{caves:81}.
\par
Following the same line of thought we investigate the possibility of
exploiting the squeezed states of light also in the case of the
estimation of the PSs covariance between the two interferometers.
Since $\hN_{k-}(\phi_k)$, $k=1,2$, can be used to estimate the PS $\phi_k$ of
the interferometer $\Int_k$, it is reasonable to evaluate the
covariance of the PSs $\mathcal{E}_\parallel \left[ \delta \phi_{1}
  \delta \phi_{2}\right]$ in Eq.~(\ref{PSs-Cov}) from
the covariance between $\hN_{1-}(\phi_1)$ and $\hN_{2-}(\phi_2)$.
We define $\widehat{C}(\phi_1,\phi_2) = \Delta
\hN_{1-}(\phi_k) ~ \Delta \hN_{2-}(\phi_k)$, with $\Delta \hN_{k-}(\phi_k)= \hN_{k-}(\phi_k)-
\mathcal{E} \left[ \hN_{k-}(\phi_k) \right]$ and $\mathcal{E} \left[ \hN_{k-}(\phi_k)
\right]=\mathcal{E}_\parallel \left[ \hN_{k-}(\phi_k)\right] =
\mathcal{E}_\perp\left[\hN_{k-}(\phi_k)\right]$, according to the properties of
the marginal distribution of $f_x(\phi_1, \phi_2)$. Thus, the expected
value of $\widehat{C}(\phi_1,\phi_2) $ in each configuration is
effectively the covariance between $\hN_{1-}(\phi_k)$ and $\hN_{2-}(\phi_k)$, i.e.,
$\mathcal{E}_x \left[\widehat{C}(\phi_1,\phi_2) \right]= \mathcal{E}_x
\left[ \hN_{1-}(\phi_k) ~\hN_{2-}(\phi_k) \right] - \mathcal{E} \left[\hN_{1-}(\phi_k)\right]
~\mathcal{E} \left[ \hN_{2-}(\phi_k)\right] $.

\par
We inject the state $|\psi\rangle=|\xi_1 \rangle_{a_1}
\otimes|\alpha_1 \rangle_{b_1} \otimes|\xi_2 \rangle_{a_2}
\otimes|\alpha_2 \rangle_{b_2}$ through $\mathcal{I}_{1}$ and
$\mathcal{I}_{2}$ considering symmetry properties of the states, $ \xi_{k}= \xi$, $\alpha_{k}=
\alpha$ and of the interferometers $\phi_{1,0}=\phi_{2,0}=\phi_{0}$, and setting $\theta_{\alpha,1}-\theta_{\xi,1}=\theta_{\alpha,2}-\theta_{\xi,2}=0$, $k=1,2$.

\par
According to Eq. (4) of the main text, and by substituting the input-output
relations (\ref{IOinterf}) into the definition of the observable
$\widehat{C}(\phi_1,\phi_2) = \Delta \hN_{1-}(\phi_1) ~ \Delta \hN_{2-}(\phi_2)$ the
minimum uncertainty becomes:
\begin{equation}
\label{U1sq} \mathcal{U}^{(0)}(\mu,\lambda,\phi_{0}=\pi/2)=\sqrt{2}~\frac{\lambda + \mu \left(1+2 \lambda - 2 \sqrt{\lambda
+\lambda^2}\right)}{(\lambda-\mu)^2}.
\end{equation}

\par
As a final comment, we remark that the advantage of the present scheme
is naturally limited by the amount of squeezing. In fact, by setting
$\phi=\pi/2$ in Eq.s~(\ref{IOinterf}), we observe that the
interferometer behaves as a 50:50 BS, and the measurement
of $N_{k-}$ corresponds to the quadrature measurement of the input mode
$a_k$ by the well known homodyne detection technique,
where the mode $b_k$ plays the role of an intense local oscillator ($\mu
\gg 1$). Increasing the intensity $\lambda$ of the squeezed field in the modes
$a_{k}$ means, by definition, reducing the quantum fluctuation on the
measured quadrature and thus enhancing the accuracy of PS estimation
in each interferometer.

\subsection{The effect of losses}\label{The effect of losses}

 The overall effect of losses can be modeled by means of BSs with
  a suitable transmittance. Formally, this corresponds to the
  substitution of the output modes $c$ and $d$ of
  Eq.s~(\ref{IOinterf}) with:
\begin{equation} \label{IOinterf-eta}
\begin{array}{ll}
c \to c_{\eta}= \eta^{1/2}~c+(1-\eta)^{1/2}~ v_{\mathrm{c}}  , \\
d \to d_{\eta}= \eta^{1/2}~d+(1-\eta)^{1/2}~ v_{\mathrm{d}} .
\end{array}
\end{equation}
that is the transmitted outputs of two identical BSs, both
with transmittance $\eta$, where the modes $v_{\mathrm{c}}$ and
$v_{\mathrm{d}}$ are taken in the vacuum state
$|0_{v_{\mathrm{c}}}\rangle\otimes|0_{v_{\mathrm{d}}}\rangle$.
Thereafter, the calculation of $\mathcal{U}_{\rm SQ}^{(0)}$ and
$\mathcal{U}_{\rm TWB}^{(0)}$ as a function of $\eta$ is performed in complete analogy of the lossless case (see Sec.~\ref{Derivation of Eq.(3) of the main text}
for independent single mode squeezed states, while for TWB see the specific paragraph of the main text).

\subsection{Radiation pressure noise}\label{Radiation pressure noise}

Our analysis is based on the reasonable hypothesis of small
phase-shift fluctuations. Therefore the main contribution to the noise
should come from the intrinsic photon noise in the measurement, given
by the zero-order term in Eq.~(\ref{VarC}). However, it is important to
test this assumption including, at least, the effect of the other
unavoidable contribution to the noise due to the radiation pressure
(RP).  In the context of our proposal, we demonstrate that, for
reasonable values of the involved parameters, RP contribution is
negligible.
\par
The statistical properties of the RP noise are determined only by the
fluctuation of the photon number inside the arms of the
interferometers, that is independent of the HN. In particular, the
fluctuation can be written as the sum of the two independent
contributions, namely, $\delta \phi_{k} = \delta \phi_{k,HN}+ \delta
\phi_{k,RP}$, while the global probability density is the product of
the HN density function with the RP one
$F(\phi_{1,HN},\phi_{1,HN},\phi_{1,RP}\phi_{2,RP})=f_x(\phi_{1,HN},
\phi_{2,HN}) \,g (\phi_{1,RP}, \phi_{2,RP})$, where we assume that $g$
is the same for $x=\parallel, \perp$. Since the variance and
covariance of the PSs over $F$ are:
\begin{eqnarray}
\mathcal{E}_x \left[\delta \phi_{k}^2\right]&=&\mathcal{E}_x
\left[\delta \phi_{k,HN}^2\right]+\mathcal{E} \left[\delta
  \phi_{k,RP}^2\right],\\
\mathcal{E}_x \left[\delta \phi_{1}\delta
  \phi_{2}\right]&=&\mathcal{E}_x \left[\delta \phi_{1,HN}\delta
  \phi_{2,HN}\right]+\mathcal{E} \left[\delta \phi_{1,RP}\delta
  \phi_{2,RP}\right],
\end{eqnarray}
respectively, the equation~(\ref{PSs-Cov}) can be written as:
\begin{equation}\label{HN-Cov}
  \mathcal{E}_\parallel \left[ \delta \phi_{1,HN} \delta \phi_{2,HN} \right]
  \approx
  \frac{\mathcal{E}_\parallel \left[ \widehat{C}(\phi_1 \phi_2)
      \right]-\mathcal{E}_\perp \left[ \widehat{C}(\phi_1, \phi_2)\right]}{\langle \partial_{\phi_{1},\phi_{2}}^{2}\widehat{C}(\phi_{1,0},\phi_{2,0}) \rangle }.
\end{equation}

\par
If the HN is absent or by far smaller than the other sources of
noise, as expected, the measurement uncertainty stemming from the
photon noise and the RP noise can be obtained by using Eq.~(\ref{VarC}) in
the numerator of Eq.~(\ref{U}):
\begin{equation}\label{URP}
 \mathcal{U}^{(2)}(\delta \phi_{1} \delta \phi_{2}) \approx
\sqrt{2 \frac{\mathrm{Var} \left[ \widehat{C}(\phi_{1,0},\phi_{2,0}) \right]+ \Sigma_k ~ A_{kk}~\mathcal{E} \left[\delta \phi_{k,RP}^2\right]+ A_{12}~\mathcal{E} \left[\delta \phi_{1,RP}\delta \phi_{2,RP}\right]}{\left[ \langle \partial_{\phi_{1},\phi_{2}}^{2}  \widehat{C}(\phi_{1,0}, \phi_{2,0}) \rangle \right]^2 }}.
\end{equation}
\par
 The difference in the transferred momentum to the two mirrors of
  the interferometer $\mathcal{I}_{k}$ is $\mathcal{P}_{k}= (2\hbar
  \omega/c)\, \widehat{n}_{-,k}$, where $\widehat{n}_{-,k}$ represents
  the photon numbers difference in the two arms and $\omega$ is the
  central frequency of the light. The arms length difference
  induced by the RP can be written as
  $z_{k,RP}=(\tau/2m)\mathcal{P}_{k}$, $\tau$ being the measurement
  time and $m$ the masses of the mirrors, and the corresponding PS is
  $\phi_{k,RP}=(\omega/c) z_{k,RP}$.
Therefore, the fluctuation $\delta\phi_{k,RP}$ is proportional to
$\delta\widehat{n}_{-,k}=\widehat{n}_{-,k}-\langle\widehat{n}_{-,k}\rangle$
and the variance and the covariance are:
\begin{subequations}\label{Var_RP,Cov_RP}
\begin{eqnarray}
\mathcal{E} \left[\delta \phi_{k,RP}^2\right]&\approx& \left(\frac{\hbar \omega^{2}\tau}{c^{2}m}\right)^{2} \langle\delta\widehat{n}_{-,k}^{2}\rangle,\\
\mathcal{E} \left[\delta \phi_{1,RP}\,\delta \phi_{2,RP}\right]&\approx& \left(\frac{\hbar \omega^{2}\tau}{c^{2}m}\right)^{2} \langle\delta\widehat{n}_{-,1}\,\delta\widehat{n}_{-,2}\rangle,
\end{eqnarray}
\end{subequations}
respectively. The quantum expectation values in the right-hand sides
of Eq.s~(\ref{Var_RP,Cov_RP}) can be evaluated easily by writing
$\widehat{n}_{-,k}=a^{\dag}_{k,out}a_{k,out}-b^{\dag}_{k,out}b_{k,out}$
according to the beam splitter input-output relations
$a_{k,out}=2^{-1/2}(a_{k}+b_{k})$ and
$b_{k,out}=2^{-1/2}(b_{k}-a_{k})$.
 In the case of independent injected squeezed states $|\psi\rangle=|\xi \rangle_{a_1}
\otimes|\alpha \rangle_{b_1} \otimes|\xi \rangle_{a_2}\otimes|\alpha \rangle_{b_2}$ we have:
\begin{subequations}\label{mean dn-SQ}
\begin{eqnarray}
 \langle\delta\widehat{n}_{-,k}^{2}\rangle_{SQ}&=& \lambda +\mu \left\{1+2 \lambda +2 \sqrt{\lambda  (1+\lambda )} \cos[2 (\theta_{\alpha,k}-\theta_{\xi,k}) ]\right\}\\
\langle\delta\widehat{n}_{-,1}\,\delta\widehat{n}_{-,2}\rangle_{SQ}&=&0.
\end{eqnarray}
\end{subequations}
where we have set the optimal working point
$\phi_{1,0}=\phi_{2,0}=\phi_{0}=\pi/2$. We note that here the
contribution of the covariance of the photon number difference is
null, because of the independence of the squeezed states sent in the
two interferometers, while the variance in each interferometer agrees
with the one reported in \cite{caves:81}. As in the usual treatment of
a single interferometer fed by squeezed light, the amplitude of the RP
noise varies with the squeezing parameter at the opposite of the
photon noise (if the photon noise reduces the RP noise increases or
viceversa) \cite{caves:81}.
\par
\begin{figure}[htbp]
\begin{center}
\includegraphics[width=0.8\textwidth]{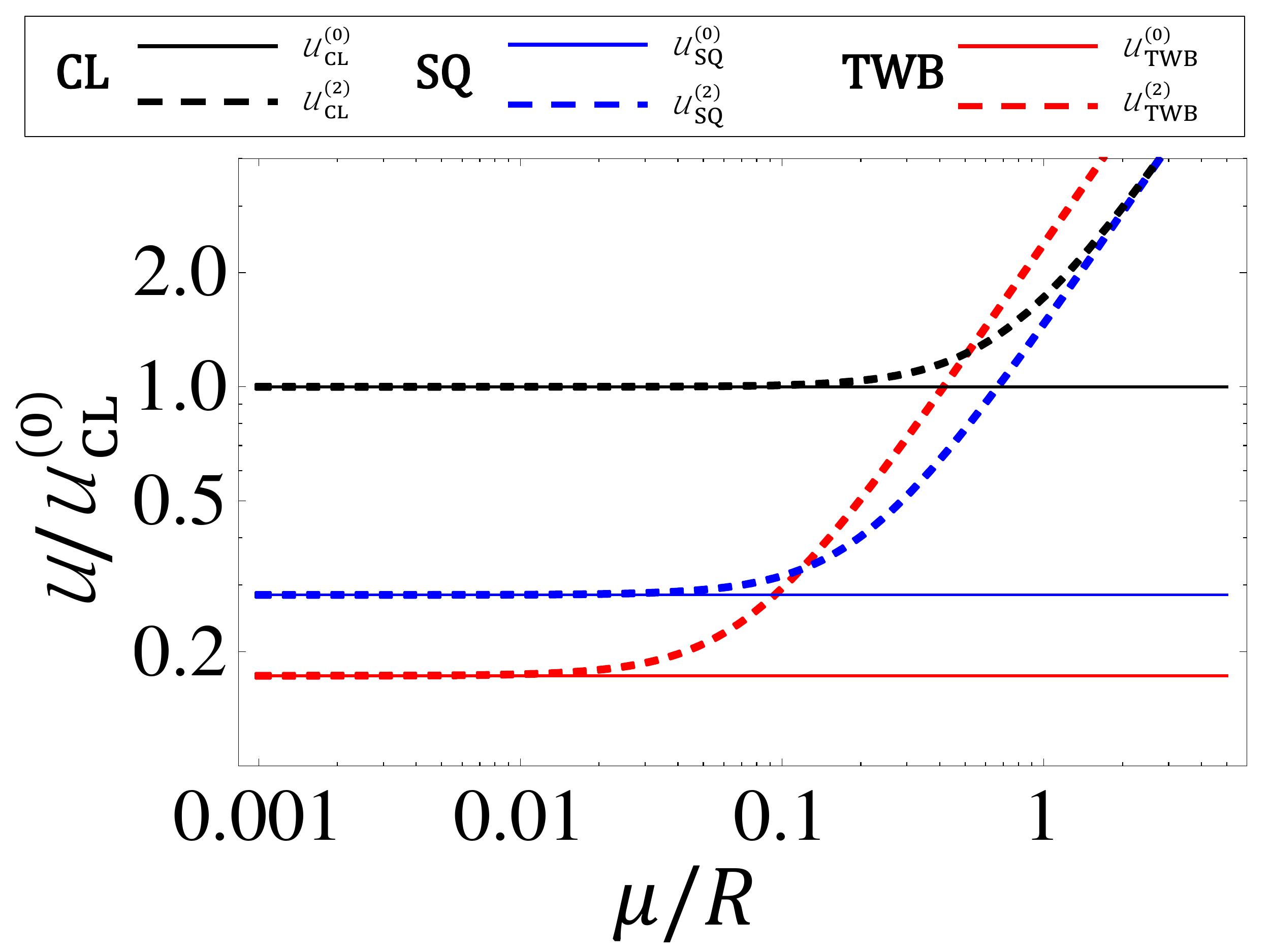}
\caption{ Uncertainty reduction normalized to the classical limit
  $\mathcal{U}_{CL}^{(0)}$, versus the normalized mean number of
  photons of the coherent fields $\mu/R$. $R=\left(\frac{\hbar
      \omega^{2}\tau}{c^{2}m}\right)^{-1}$ is the characteristic
  adimensional parameter connected with the phase-fluctuation response
  to the photon number fluctuation, according to
  Eq.~(\ref{Var_RP,Cov_RP}): one has $R\approx 8.6 \times 10^{24}$ for
  the values of $\tau,\, m$ and $\omega$ chosen in Fig.~3 of the main
  text. The solid lines represent the uncertainties only due to the
  photon noise, corresponding to the zero-order contribution [see
  Eq.~(\ref{URP})], while the dashed lines represent the second-order
  uncertainties including the RP noise. The twin beams and independent
  squeezed beams intensities are $\lambda_{1}=\lambda_{2}=\lambda=0.5$
  and the overall transmission-detection efficiency $\eta=0.98$}
\label{fig_U-vs-mu}
\end{center}
\end{figure}

When the expectation values in Eq.s~(\ref{Var_RP,Cov_RP}) are
evaluated for the TWB case, i.e. for the state $|\psi\rangle=|{\rm
  TWB}\rangle\rangle_{a_1,a_2}\otimes|\alpha\rangle_{b_1}\otimes|\alpha\rangle_{b_2}$,
we obtain:
\begin{subequations}\label{mean dn-TWB}
\begin{eqnarray}
 \langle\delta\widehat{n}_{-,k}^{2}\rangle_{TWB}&=& \lambda + \mu (1 + 2 \lambda)\\
\langle\delta\widehat{n}_{-,1}\,\delta\widehat{n}_{-,2}\rangle_{TWB}&=&2 \mu \sqrt{\lambda (1 + \lambda)} \cos[2 (\theta_{\alpha,k}-\theta_{\zeta,k})].
\end{eqnarray}
\end{subequations}
where we have set the optimal working point
$\phi_{1,0}=\phi_{2,0}=\phi_{0}=0$. Comparing Eq.s (\ref{mean dn-SQ})
and Eq.s~(\ref{mean dn-TWB}) one notes that the single interferometer
contribution is of the same order for both ``SQ'' and ``TWB''
configurations, while the covariance of the ``TWB'' is non null and
basically similar to the variance of the RP of the single
interferometer (if $\theta_{\alpha,k}-\theta_{\zeta,k}=0$, as
considered through all the paper). Even if the effect of the RP noise
on the final measurement is given by Eq.~(\ref{URP}), and requires the
non trivial calculation of the coefficients $A_{kj}$, this suggests
that the RP could be more effective in the case of TWB. We do not
  report here the lengthy calculation which leads to a cumbersome result;
  nevertheless, we show in Fig.~\ref{fig_U-vs-mu} the behavior of the
  uncertainties $\mathcal{U}$ as functions of the mean number of
  photons of the coherent state $\mu$. It is worth noting that RP
  contribution is negligible for reasonable value of $\mu$, while for
  non-realistic higher value of $\mu$ where RP becomes relevant, TWB is a
  bit more penalized comparing with the case of independent squeezed
  states.
\par
Summarizing, the advantage of using twin beams is not affected substantially for
reasonable values of the parameters, as shown also in Fig.~3 of the main
text, where the uncertainty $\mathcal{U}^{(2)}(\delta \phi_{1} \delta
\phi_{2})$ of Eq.~(\ref{URP}) is plotted as a function of the detection
efficiency (see Sec.~\ref{The effect of losses}).

{\bf Acknowledgements}

The research leading to these results has received funding from the EU FP7 under grant agreement n.
308803 (BRISQ2),   Fondazione San
Paolo and MIUR (FIRB ``LiCHIS'' - RBFR10YQ3H and Progetto Premiale  ``Oltre i limiti classici di misura''), NATO grant EAP-SFPP98439.


\begin{thebibliography}{30}
\bibitem{footnote1} An other useful interpretation of this quantity is related to the signal-to-noise ratio (SNR) of the measurement.
In general $SNR\equiv S/\sqrt{\hbox{Var}(S)}=(K\, E)/\sqrt{\hbox{Var}(S)}$ where $S$ is the signal ($S=\mathcal{E}_\parallel \left[ \widehat{C}\right]-\mathcal{E}_\perp \left[ \widehat{C}\right]$ for us)
 that in a linear approximation is related by a certain constant $K$ ($Kk$ is the denominator of Eq. (\ref{cov:PS})) to the quantity under estimation $E$,
 in our case is the phase covariance $E=\mathcal{E}_\parallel\left [ \delta \phi_{1} \delta \phi_{2}\right]$). The minimum resolvable value of $E$, i.e. the one that allows to reach a SNR larger than 1, is $E_{min}=\sqrt{\hbox{Var}(S)}/K$, i.e.
 exactly $\mathcal{U}$.
\end{thebibliography}
\end{document}